\documentclass[11pt]{article}
\usepackage{amsmath,amsthm,latexsym,amssymb,amsfonts,epsfig}

\newtheorem{lemm}{Lemma}[section]

\newtheorem{defi}[lemm]{Definition}

\newcommand{\skripta}{\mathfrak{A}}        
\newcommand{\skriptw}{\mathfrak{W}}

\newcommand{\one}{\text{\bf 1}}               

\newcommand{\gcon}{\overline{\mathcal{A}}}    
\newcommand{\con}{\mathcal{A}}                
\newcommand{\comm}[2]{\left[#1\,,\,#2\right]} 
 
\newcommand{\scpr}[2]{\left\langle#1\,,\, #2 \right\rangle} 

\newcommand{\R}{\mathbb{R}}                  
            
\DeclareMathOperator{\porder}{\mathcal{P}}   
\DeclareMathOperator{\cyl}{Cyl}              

\DeclareMathOperator{\sutwo}{SU(2)}          

\makeatletter
\@addtoreset{equation}{section}
\makeatother

\newtheorem{Theorem}{Theorem}[section]

\newtheorem{Lemma}{Lemma}[section]

\def\be{\begin{equation}}
\def\ee{\end{equation}}
\def\ba{\begin{eqnarray}}
\def\ea{\end{eqnarray}}
\def\a{{\cal A}}
\def\ab{\overline{\a}}

\def\Nl{{\mathchoice
{\setbox0=\hbox{$\displaystyle\rm N$}\hbox{\hbox to0pt
{\kern0.4\wd0\vrule height0.9\ht0\hss}\box0}}
{\setbox0=\hbox{$\textstyle\rm N$}\hbox{\hbox to0pt
{\kern0.4\wd0\vrule height0.9\ht0\hss}\box0}}
{\setbox0=\hbox{$\scriptstyle\rm N$}\hbox{\hbox to0pt
{\kern0.4\wd0\vrule height0.9\ht0\hss}\box0}}
{\setbox0=\hbox{$\scriptscriptstyle\rm N$}\hbox{\hbox to0pt
{\kern0.4\wd0\vrule height0.9\ht0\hss}\box0}}}}
\def\Zl{{\mathchoice
{\setbox0=\hbox{$\displaystyle\rm Z$}\hbox{\hbox to0pt
{\kern0.4\wd0\vrule height0.9\ht0\hss}\box0}}
{\setbox0=\hbox{$\textstyle\rm Z$}\hbox{\hbox to0pt
{\kern0.4\wd0\vrule height0.9\ht0\hss}\box0}}
{\setbox0=\hbox{$\scriptstyle\rm Z$}\hbox{\hbox to0pt
{\kern0.4\wd0\vrule height0.9\ht0\hss}\box0}}
{\setbox0=\hbox{$\scriptscriptstyle\rm Z$}\hbox{\hbox to0pt
{\kern0.4\wd0\vrule height0.9\ht0\hss}\box0}}}}
\def\Ql{{\mathchoice
{\setbox0=\hbox{$\displaystyle\rm Q$}\hbox{\hbox to0pt
{\kern0.4\wd0\vrule height0.9\ht0\hss}\box0}}
{\setbox0=\hbox{$\textstyle\rm Q$}\hbox{\hbox to0pt
{\kern0.4\wd0\vrule height0.9\ht0\hss}\box0}}
{\setbox0=\hbox{$\scriptstyle\rm Q$}\hbox{\hbox to0pt
{\kern0.4\wd0\vrule height0.9\ht0\hss}\box0}}
{\setbox0=\hbox{$\scriptscriptstyle\rm Q$}\hbox{\hbox to0pt
{\kern0.4\wd0\vrule height0.9\ht0\hss}\box0}}}}
\def\Rl{{\mathchoice
{\setbox0=\hbox{$\displaystyle\rm R$}\hbox{\hbox to0pt
{\kern0.4\wd0\vrule height0.9\ht0\hss}\box0}}
{\setbox0=\hbox{$\textstyle\rm R$}\hbox{\hbox to0pt
{\kern0.4\wd0\vrule height0.9\ht0\hss}\box0}}
{\setbox0=\hbox{$\scriptstyle\rm R$}\hbox{\hbox to0pt
{\kern0.4\wd0\vrule height0.9\ht0\hss}\box0}}
{\setbox0=\hbox{$\scriptscriptstyle\rm R$}\hbox{\hbox to0pt
{\kern0.4\wd0\vrule height0.9\ht0\hss}\box0}}}}
\def\Cl{{\mathchoice
{\setbox0=\hbox{$\displaystyle\rm C$}\hbox{\hbox to0pt
{\kern0.4\wd0\vrule height0.9\ht0\hss}\box0}}
{\setbox0=\hbox{$\textstyle\rm C$}\hbox{\hbox to0pt
{\kern0.4\wd0\vrule height0.9\ht0\hss}\box0}}
{\setbox0=\hbox{$\scriptstyle\rm C$}\hbox{\hbox to0pt
{\kern0.4\wd0\vrule height0.9\ht0\hss}\box0}}
{\setbox0=\hbox{$\scriptscriptstyle\rm C$}\hbox{\hbox to0pt
{\kern0.4\wd0\vrule height0.9\ht0\hss}\box0}}}}
\def\Hl{{\mathchoice
{\setbox0=\hbox{$\displaystyle\rm H$}\hbox{\hbox to0pt
{\kern0.4\wd0\vrule height0.9\ht0\hss}\box0}}
{\setbox0=\hbox{$\textstyle\rm H$}\hbox{\hbox to0pt
{\kern0.4\wd0\vrule height0.9\ht0\hss}\box0}}
{\setbox0=\hbox{$\scriptstyle\rm H$}\hbox{\hbox to0pt
{\kern0.4\wd0\vrule height0.9\ht0\hss}\box0}}
{\setbox0=\hbox{$\scriptscriptstyle\rm H$}\hbox{\hbox to0pt
{\kern0.4\wd0\vrule height0.9\ht0\hss}\box0}}}}
\def\Ol{{\mathchoice
{\setbox0=\hbox{$\displaystyle\rm O$}\hbox{\hbox to0pt
{\kern0.4\wd0\vrule height0.9\ht0\hss}\box0}}
{\setbox0=\hbox{$\textstyle\rm O$}\hbox{\hbox to0pt
{\kern0.4\wd0\vrule height0.9\ht0\hss}\box0}}
{\setbox0=\hbox{$\scriptstyle\rm O$}\hbox{\hbox to0pt
{\kern0.4\wd0\vrule height0.9\ht0\hss}\box0}}
{\setbox0=\hbox{$\scriptscriptstyle\rm O$}\hbox{\hbox to0pt
{\kern0.4\wd0\vrule height0.9\ht0\hss}\box0}}}}

\title{Irreducibility of the Ashtekar -- Isham -- Lewandowski 
Representation}
\author{
Hanno Sahlmann\thanks{hanno@gravity.psu.edu},\\
Center for Gravitational Physics and Geometry, \\
The Pennsylvania State University, University Park, PA, USA\\
\\
Thomas Thiemann\thanks{thiemann@aei-potsdam.mpg.de,
tthiemann@perimeterinstitute.ca}\thanks{New Address:
The Perimeter Institute for Theoretical Physics
and Waterloo University, Waterloo, Ontario, Canada},\\
MPI f\"ur Gravitationsphysik, Albert-Einstein-Institut, \\
Am M\"uhlenberg 1, 14476 Golm near Potsdam, Germany
}
\date{{\small Preprint AEI-2003-034, PI-2003-002, CGPG-03/3-3}}

\begin{document}

\maketitle

\begin{abstract}
Much of the work in loop quantum gravity and quantum geometry rests on a 
mathematically rigorous integration theory on spaces of distributional
connections. Most notably, a diffeomorphism invariant representation
of the algebra of basic observables of the theory, the
Ashtekar-Isham-Lewandowski representation, has been
constructed. Recently, several uniqueness results for this
representation have been worked out.  
In the present article, we contribute to these efforts by showing that
the AIL-representation is irreducible, provided it is viewed as the
representation of a certain C$^*$-algebra which is very similar to the
Weyl algebra used in the canonical quantization of free quantum field
theories.
\end{abstract}

\section{Introduction}
\label{s1}

Canonical, background independent quantum field theories of connections 
\cite{1} play a fundamental role in the program of canonical 
quantization of general relativity (including all types of matter),
sometimes called loop quantum gravity or quantum general relativity
(for a review geared to mathematical physicists see \cite{2}). The 
classical canonical theory can be formulated in terms of smooth 
connections $A$ on principal $G-$bundles over a $D-$dimensional spatial 
manifold 
$\Sigma$ for a compact gauge group $G$ and smooth sections of an 
associated (under the adjoint representation) vector bundle of 
Lie$(G)-$valued vector densities $E$ of weight one. The pair 
$(A,E)$ coordinatizes an infinite dimensional symplectic manifold 
$({\cal M},\sigma)$ whose (strong) symplectic structure $s$ is such
that $A$ and $E$ are canonically conjugate.

In order to quantize $({\cal M},s)$, it is necessary to smear the 
fields $A,E$. This has to be done in such a way that the smearing 
interacts well with two fundamental automorphisms of the principal 
$G-$ bundle, namely the vertical automorphisms formed by $G-$gauge 
transformations and the horizontal automorphisms formed by Diff$(\Sigma)$ 
diffeomorphisms. These requirements naturally
lead to holonomies and electric fluxes, that is, exponentiated 
(path-ordered) 
smearings of the connection over $1-$dimensional submanifolds $e$ of $\Sigma$ 
as well as smearings of the electric field over $(D-1)-$dimensional 
submanifolds $S$,
\begin{equation*}
  h_e[A]=\porder \exp i \int_e A,\qquad E_{S,n}[E]=\int_S *E_i n^i.
\end{equation*}
These functions on $\cal M$ generate a closed Poisson $\ast$-algebra 
$\cal P$ and separate the points of $\cal M$. They do not depend on a
choice of coordinates nor on a background metric. Therefore, 
diffeomorphisms and gauge transformations act on these variables in a
remarkably simple way.

Quantization now means to promote $\cal P$ to an abstract
$\ast$-algebra $\skripta$
and to look for its representations. Definitions of $\skripta$ have 
been given in \cite{10,10a,10b}, and we will review what
we need in the next section.
Remarkably, $\skripta$ admits a simple and mathematically elegant 
representation, the Ashtekar-Isham-Lewandowski representation 
$({\cal  H}_0,\pi_0)$ 
\cite{6,7}. An important feature of this representation is that 
it contains a cyclic vector that is invariant under the action of both
diffeomorphisms and gauge transformations. Therefore it is a good
starting point to tackle the implementation of the Gauss- and the
diffeomorphism constraint of gravity \cite{1}. 

In the present work we will add another item to the list of desirable
properties that the AIL representation possesses: We will show that it
is irreducible in a specific sense. 
The first thing we have to point out in this context 
is that $\skripta$ is not an
algebra of bounded operators. Consequently, to define a notion of 
irreducibility one has to worry about domain questions, and
definitions as well as proofs become rather cumbersome. Maybe it is
due to these technical difficulties that up to now, little has been said
concerning irreducibility of the AIL representation.  
To the best of our knowledge, the only result in this direction is 
that the algebra $\skripta$ allows us to map between any 
two vectors $f,f'\in {\cal D}$ where ${\cal D}$ is a dense subset of 
${\cal H}_0$ \cite{2}. 

In this situation, it is worthwhile to note that there is a strong 
analogy between the AIL representation of $\skripta$ and the Schr\"odinger 
representation of the Heisenberg algebra in quantum mechanics. 
In both representations, the representation spaces are roughly speaking 
$L_2$ spaces over the configuration space, the configuration variables
act by multiplication and the momenta by derivations. The Heisenberg
algebra is again an algebra of unbounded operators, which makes the
definition of irreducibility difficult. Moreover it is dubious that
its Schr\"odinger representation can be irreducible in any sense,
since for example the subspaces generated by functions which vanish on
fixed open sets are invariant under action with multiplication
operators and differentiation (if defined). 
However, the Schr\"odinger representation of the Heisenberg algebra
can be obtained from the Schr\"odinger representation of the
corresponding Weyl algebra, and it is \textit{this} representation that is
irreducible. In fact, von Neumann's famous uniqueness result states
that it is the \textit{only} irreducible, strongly continuous
representation of the Weyl algebra.  

In the light of this analogy, it seems worthwhile to investigate
whether the AIL representation derives from a representation of an
algebra of bounded operators which is irreducible. In fact, in
\cite{10a} we introduced an algebra $\skriptw$ that contains the
unitary one-parameter groups generated by the fluxes $E_{S,f}$
instead of the fluxes themselves. This algebra of \textit{bounded}
operators turns out to be closely  
analogous to the Weyl algebra used in the quantization of free fields,
and the AIL representation of $\skripta$ can be recovered from a
strongly continuous representation of $\skriptw$ (which, in the following,
will also be called ``AIL representation''). The main result of \cite{10a}
was that requiring diffeomorphism invariance, strong continuity, and certain
technical conditions uniquely singles out the AIL-representation, very
much in analogy to von Neumann's theorem. In the present
article, we will carry this analogy further by showing that the AIL
representation of $\skriptw$ is indeed irreducible.\\
\\
The article is organized as follows:\\
\\
In section \ref{s2} we recall from \cite{1,6,7,9} the essentials of the 
classical formulation 
of canonical, background independent theories of connections, that is,
the symplectic manifold $({\cal M},\sigma)$ and the 
classical Poisson $\ast$-algebra generated by holonomies and 
electric fluxes. We then recall from \cite{10,10a} the definition of 
the corresponding abstract algebra $\skripta$ and its companion $\skriptw$.\\
In section \ref{s3} we prove irreducibility of the AIL representation
of $\skriptw$.\\
In section \ref{s4} we finish with some conclusions.\\
\\
\section{Preliminaries}
\label{s2}
This section serves to review the definitions of the algebras
$\skripta$ and $\skriptw$. As most of this has been treated in detail
elsewhere, we will just give an overview and refer to the appropriate
literature for details.

Let $G$ be a a compact, connected  Lie group. For convenience, fix a basis 
$\tau_i$ of the corresponding Lie algebra. The resulting indices are
dragged with the Cartan-Killing metric on $G$,
although for simplicity we will not write this explicitely.  
Let $\Sigma$ be an analytic, connected and orientable $D-$dimensional
and $P$ a principal $G$-bundle over $\Sigma$. 
The smooth connections in $P$ form the
classical configuration space $\con$, the corresponding momenta can be
identified with sections in a vector bundle $E_P$ associated to $P$ under the 
adjoint representation, whose typical fiber is a Lie$(G)-$valued 
$(D-1)-$form on $P$. In a local trivialization, the Poisson brackets
read 
\begin{equation}
\label{eq1}
  \{A^i_a(x),A^{i'}_{a'}(x')\}=\{E^a_i(x),E^{a'}_{i'}(x')\}=0,\qquad
\{A^i_a(x),E^{a'}_{i'}(x')\}=\delta_a^{a'}\delta_{i'}^{i}\delta(x,x').
\end{equation}
In a concrete gauge field theory the right hand side of the last
equation will be multiplied 
by a constant which depends on the coupling constant of the theory.
In order not to clutter our formulae we will assume that $A$ and $E$ 
respectively have dimension cm$^{-1}$ and cm$^{-(D-1)}$ respectively.

As pointed out in the introduction, it turned out to be very fruitful
to go over to certain functionals of $A$ and $E$. For the connection,
the functionals are chosen to be the parallel transports along
analytical paths $e$ in $\Sigma$,  
\begin{equation*}
h_e[A]=\porder\exp\left[i\int_e A_a ds^a \right].
\end{equation*}
In fact, it turns out to be convenient to consider functions of $A$ which are 
slightly more general.
\begin{defi}
\label{def1}
    A \textit{graph} in $\Sigma$ is a collection of analytic, oriented curves 
    in $\Sigma$ which intersect each other at most in their 
    endpoints. Given a graph $\gamma$, the set of its constituting
    curves (``edges'') will be denoted by $E(\gamma)$.

    A function $f$ depending on connections $A$ on $\Sigma$ just in
    terms of their holonomies along the edges of a graph, i.e. 
    \begin{equation*}
      f[A]\equiv f(h_{e_1}[A],h_{e_2}[A],\ldots,h_{e_n}[A] ), \qquad
      e_1,e_2,\ldots, e_n\quad\text{ edges of some }\gamma, 
    \end{equation*}
    where $f(g_1,\ldots, e_n)$ viewed as a function on $\sutwo^n$ is
    \textit{continuous}, will be called \textit{cylindrical}. 
\end{defi}
It turns out that the set of cylindrical functions can be equipped with 
a norm (essentially the sup-norm for functions on SU(2)$^n$) such that 
its closure (denoted by $\cyl$) with respect to that norm is a
commutative C$^*$-algebra. We will not spell out the details of this
construction but refer the reader to the presentations
\cite{7,9}. We note furthermore that by changing the word
``continuous'' 
in the above definition to ``$n$ times differentiable'', we can define
subsets $\cyl^n$ of $\cyl$ and, most importantly for us, 
\begin{equation*}
  \cyl^\infty:= \bigcap_n\cyl^n, 
\end{equation*}
the space of smooth cylindrical functions. 

By Gelfand's theory, $\cyl$ can be identified with the continuous
function on a compact Hausdorff space $\gcon$. The inclusion of the
functions on $\con$ defined in Definition \ref{def1} into the
continuous functions on $\gcon$ is afforded by the fact that $\gcon$ is
a projective limit: Each graph $\gamma$ defines a projection 
$p_\gamma:\gcon\longrightarrow\gcon_\gamma$, where $\gcon_\gamma$ is
diffeomorphic to $G^N$, $N$ being the number of edges of 
$\gamma$.\footnote{The diffeomorphism is not unique but roughly
  speaking depends on the choice of a gauge. But we can ignore this
  fact since the algebraic structures one obtains in the end are not 
  affected by this choice.}

The density weight of $E$ on the other hand is such that, using an
additional real (co-)vector field $n^i$, it can be naturally integrated
over oriented surfaces $D-1$ dimensional submanifolds $S$ to form a quantity
\begin{equation*}
E_{S,n}= \int_S E^a_i n^i \epsilon_{abc}\,dx^b\,dx^c
\end{equation*}
analogous to the electric flux through $S$. In the following we will
only consider analytic submanifolds $S$ to avoid certain pathologies
in the algebra relations to be defined below. 

It has been shown in \cite{19} that the algebra generated by $\cyl$ and
the fluxes $\{E_{S,n}\}$ can be given the structure of a Lie algebra
which derives in a precise sense from the Poisson relations 
\eqref{eq1}. To spell out this structure, we will first define the
action of certain derivations $Y_{S,n}$ on $\cyl$, where again $S$ is
an analytical surface and $n$ a vectorfield in $E_P|_S$. 
Let $f$ be a smooth function cylindrical on $\gamma$ and assume
without loss of generality that all transversal intersections of
$\gamma$ and $S$ are in vertices of $\gamma$. Then one defines   
\begin{equation*}
  Y_{S,n}[f]:= \frac{1}{2}\sum_{p\in S\cap\gamma}
  \sum_{e_{p}} \sigma(e_{p},S)n_i(p)Y^i_{e_{p}}[f],
\end{equation*}
where the second sum is over the edges of $\gamma$ adjacent to $p$, 
\begin{equation*}
    \sigma(e_{p},S)=\begin{cases}1&\text{ if $e_p$ lies above $S$}\\
0&\text{ if $e_p\cap S=\emptyset$ or $e_p\cap S=e_p$}\\ 
-1&\text{ if $e_p$ lies below $S$}\end{cases},
\end{equation*}
and $Y^i_{e_{p}}$ is the $i$th left-invariant (right-invariant) 
vector field on SU(2) acting on the argument of $f$ corresponding to 
the holonomy $h_{e_{p}}$ if $e_{p}$ is pointing away from (towards) $S$.

A Lie product between elements of $\cyl^\infty$ and fluxes
$\{E_{S,n}\}$ can now be defined as
\begin{equation}
\label{eq2}
  \{f,E_{S,n}\}=Y_{S,n}[f]. 
\end{equation}
A bit surprisingly at first sight, it turns out that Lie products
between elements of $\{E_{S,n}\}$ can not, in general vanish. However,
their Lie products with cylindrical functions are
completely determined by \eqref{eq2} together with the requirement
that the Jacobi identity has to hold.\footnote{The fact that the fluxes 
among themselves do not commute can be traced
back to the fact that going over from \eqref{eq1} to \eqref{eq2}
involves a nontrivial limiting procedure. For more information on this
point see \cite{19}.} 
For example one finds
\begin{equation*}
  \{f,\{E_{S,n},E_{S',n'}\}\}=\comm{X_{S',n'}}{X_{S,n}}[f].
\end{equation*}
We are now in a position to define the algebra $\skripta$:
\begin{defi}
Let $\skripta$ be the algebra generated by $\cyl$ together with
symbols $\{E_{S,n}\}$, divided by the commutation relations 
\begin{equation*}
\comm{E_{S,n}}{f}=\frac{1}{i}Y_{S,n}[f].
\end{equation*}
Equip $\skripta$ with an involution by defining 
\begin{equation*}
  f^*:=\overline{f},\qquad E_{S,n}^*:=E_{S,n}.
\end{equation*}
\end{defi}
The general representation theory of the algebra $\skripta$ 
gets complicated due to 
the fact that the $\{E_{S,n}\}$ will be represented by unbounded 
operators, so that domain questions will arise. In \cite{10a} we
proposed to circumvent these difficulties by 
passing to exponentials of $i$ times the fluxes, thereby obtaining an
algebra $\skriptw$ analogous to the Weyl algebra used in the
quantization of free field theories: 
\begin{defi} \label{def2.4} ~~~~\\
Let $\skriptw$ be the algebra generated by $\cyl$ together with
symbols $W^f_t(S)$, $t\in\R$, divided by the relations 
\begin{gather*}
 W_{t_1+t_2}^n(S)=W_{t_1}^n(S)W_{t_2}^n(S), \qquad
W^n_0(S)=\one,\\ 
W_t^n(S)\, f({h_e}_{e\in E(\gamma)})\, W_{-t}^n(S)
=f(\{e^{t \sigma(S,e) n^j(b(e)) \tau_j} h_e\}_{e\in E(\gamma)})
\end{gather*}
where $f\in\cyl$ is supposed to be cylindrical on $\gamma$, and $b(e)$
denotes the starting point of $e$. Equip
$\skriptw$ with an involution by defining
\begin{equation*}
  f^*:=\overline{f},\qquad (W^n_t(S))^*:=W^n_{-t}(S).
\end{equation*}
\end{defi}
We note that these relations follow from formally identifying
$W_t^n(S)$ with $\exp(itE_{S,n})$ and using the relations
\eqref{eq2}. 

Finally, we want to describe the AIL representation of $\skripta$, 
show that it defines a representation of $\skriptw$ as well, and
notice that it can consequently be used to define a C$^*$-norm on
$\skriptw$. 

The representation space of the AIL representation is given by 
${\cal H}_0=L_2(\gcon,d\mu_0)$ where $\gcon$ is the spectrum of the 
$C^\ast-$subalgebra of $\skripta$ (and $\skriptw$) given by $\cyl$ and 
$\mu_0$ is a 
regular Borel probability measure on $\gcon$ consistently defined by
\be \label{2.18}
\mu_0(p_\gamma^\ast f_\gamma)=\int_{G^{|E(\gamma)|}} 
\prod_{e\in E(\gamma)} d\mu_H(h_e) f_\gamma(\{h_e\}_{e\in E(\gamma)})
\ee
for measurable $f_\gamma$ and extended by $\sigma-$additivity. 
Then
\be \label{2.19}
\pi_0(f) \psi[A]=f[A]\psi[A],\qquad 
\pi_0(E_{S,n}) f= \frac{1}{i}Y_{S,n}[f]  
\ee
(where $\psi \in {\cal H}_0$, $f\in \cyl^\infty\hookrightarrow{\cal
  H}_0$), define a representation of $\skripta$. Note especially that
the $\pi_0(E_{S,n})$ as defined above can be closed to selfadjoint
operators. 
This representation of $\skripta$ also defines a representation of
$\skriptw$, which, in abuse of notation, we will also denote by
$\pi_0$, via
\begin{equation*}
  \pi_0(W^n_t(S))=e^{it\pi_0(E_{S,n})}=e^{tY_{S,n}}.
\end{equation*}
It follows from the left invariance of the Haar measure that 
$\pi_0(W^f_t(S))$ are unitary operators as they should be.

Note also that the continuous functions $C^0(\ab)(\simeq\cyl)$ are 
dense in ${\cal H}_0$ because 
${\cal H}_0$ is the GNS Hilbert space induced by the positive linear 
functional $\omega_0$ on $\cyl$ defined by $\omega_0(f)=\mu_0(f)$
where $\Omega_0=1$ is the cyclic GNS vector.

Finally we equip the algebra $\skriptw\,$ with a C$^\ast$ structure.
Recall \cite{20} that if a Banach algebra admits a C$^\ast$ norm at 
all then it is unique and determined purely algebraically by
$||a||=\sqrt{a^\ast a}$ where $\rho$ denotes the spectral radius of 
$a\in\skriptw$. Since the operator norm in a representation $\pi$
of $\skriptw$ on a Hilbert space $\cal H$ {\it does} define a C$^\ast-$
norm through $||a||:=||\pi(a)||_{{\cal H}}$ we just need to find a 
representation of $\skriptw$ (and complete it in the corresponding 
operator norm). However, the Ashtekar-Lewandowski 
Hilbert space ${\cal H}_0$ {\it is} a representation space for a 
representation $\pi_0$, hence a C$^\ast-$norm exists. Let us compute
it explicitely:
As remarked earlier, the $\pi_0(W^n_t(S)$ are 
unitary operators, thus 
\ba \label{2.20}
||f||_{\skriptw}&=&||\pi_0(f)||_{{\cal B}({\cal H}_0)}
=\sup_{||\psi||=1} ||f\psi||_{{\cal H}_0}=\sup_{a\in \a} |f(A)|
\nonumber\\
||W^n_t(S)||_{\skriptw}&=&||\pi_0(W^n_t(S))||_{{\cal B}({\cal H}_0)}
=\sup_{||\psi||=1} ||W^n_t(S)\psi||_{{\cal H}_0}=1
\ea
where $\cal B$ denotes the bounded operators on a Hilbert space. The 
$C^\ast-$norm of any other element of $\skriptw$ can be computed by 
using the commutation relations and the inner product on ${\cal H}_0$.

Certainly other completions of $\skriptw$ might exist, but this is of
no concern here as we will
not make essential use of the C$^\ast-$norm in the present paper.

Finally, let us recall that the Hilbert space ${\cal H}_0$ 
has an orthonormal basis given by spin
network functions \cite{21}. These are particular cylindrical functions
labeled by a spin network 
$s=(\gamma,\{\pi_e\},\{m_e\},\{n_e\})_{e\in E(\gamma)}$ defined by
\be \label{2.21}
T_s(A)=\prod_{e\in E(\gamma)}\;
\{\sqrt{d_{\pi_e}}\;[\pi_e(h_e)]_{m_e n_e}\}
\ee
where $\gamma$ denotes a graph,
$\pi$ denotes an irreducible representation of $G$ (one fixed 
representative from each equivalence class), $d_\pi$ its dimension and 
$[\pi(h)]_{mn},\;m,n=1,..,d_\pi$ denote the matrix elements of $\pi(h)$
for $h\in G$. We write $\gamma(s)$ when $s=(\gamma,.,.,.)$.\\
\\
This concludes our exposition about the C$^\ast-$algebra $\skriptw$
and its representation $\pi_0$ on ${\cal H}_0=L_2(\ab,d\mu_0)$.

\section{Irreducibility Proof}
\label{s3}

\begin{Theorem} \label{th3.1} ~~~~\\
The Ashtekar -- Isham -- Lewandowski representation $\pi_0$ of the 
algebra $\skriptw$ on ${\cal H}_0$ is irreducible.
\end{Theorem}
Before we prove the theorem, we first need two preparational results.
Let $\gamma$ be a graph. Split each edge $e\in E(\gamma)$ into two 
halves $e=e'_1\circ (e'_2)^{-1}$ and replace the $e$'s by the $e'_1,e'_2$.
This leaves the range of $\gamma$ invariant but changes the set of 
edges in such a way that each edge is outgoing from the vertex $b(e')=v\in 
V(\gamma)$ (notice that by a vertex we mean a point in $\gamma$ which 
is not the interior point of an analytic curve so that the break points 
$e'_1\cap e'_2$ do not count as vertices). We call a graph refined in this 
way a {\it standard graph}. Every cylindrical function over a graph is 
also cylindrical over its associated standard graph so there is no loss
of generality in sticking with standard graphs in what follows.

With this understanding, the following statement holds.
\begin{Lemma} \label{la3.1} ~~~~\\
Let $\gamma$ be a standard graph. Assign to each $e\in E(\gamma)$ a vector 
$t_e=(t_e^j)_{j=1}^{\dim(G)}$ and collect them into a label 
$t_\gamma=(t_e)_{e\in E(\gamma)}$. 

Then there exists a vector field $Y(t_\gamma,\gamma)$
in the Lie algebra of the flux vector fields $Y_{S,f}$ such that for 
any cylindrical function $f=p_\gamma^\ast f_\gamma$ over $\gamma$ we have
\be \label{3.1}
Y_\gamma(t_\gamma) p_\gamma^\ast f_\gamma=
p_\gamma^\ast \sum_{e\in E(\gamma)} t_e^j R^e_j f_\gamma
\ee
\end{Lemma}
Proof of lemma \ref{th3.1}:\\
Any compact connected Lie group $G$ has the structure 
$G/Z=A\times S$ where $Z$ is a discrete central subgroup, $A$ is an 
Abelean Lie group group and $S$ is a semisimple Lie group.\\
We will first construct an appropriate vector field 
$Y^j_e$ for each $j$ and each $e\in E(\gamma)$. The construction is 
somewhat different for the Abelean and non-Abelean generators respectively
so that we distinguish the two cases.\\
{\it Abelean Factor}\\
Let $j$ label only Abelean generators for this paragraph. 
Consider any $e\in E(\gamma)$ and take any surface $S_e$ which intersects
$\gamma$ only in an interior point of $e$ and such that the orientation of 
$S_e$ agrees with that of $e_2$ where $e=e_1\circ e_2,\;e_1\cap 
e_2=S_e\cap \gamma$. Then for any cylindrical function $f=p_\gamma^\ast 
f_\gamma$ we have 
\be \label{3.2}
Y_j(S_e) p_\gamma^\ast f_\gamma= p_\gamma^\ast [R^j_{e_2}-R^j_{e_1}] 
f_\gamma
\ee
Due to gauge invariance $[R^j_{e_1}+R^j_{e_2}] f_\gamma=0$, thus
\be \label{3.3}
Y^j_e p_\gamma^\ast f_\gamma=\frac{1}{2}
Y_j(S_e) p_\gamma^\ast f_\gamma
\ee
is an appropriate choice.\\
\\
{\it Non-Abelean Factor}\\
Let $j$ label only non-Abelean generators for this paragraph. 
Given $\gamma$ select a vertex $v$ and one $e\in E(\gamma)$ with 
$b(e)=v$. 
We claim that there exists an analytic surface $S_{v,e}$ through 
$v$ such that $s_e\subset S_{v,e}=\gamma\cap S_{v,e}$ for some 
beginning segment $s_e$ of $e$ but such that any 
other $e'\in E(\gamma)$ 
is transversal to $S_{v,e}$. The analytic surface $S$ is completely 
determined by its germ $[S]_v$, that is, the Taylor coefficients in the 
expansion of its parameterization 
\be \label{3.4}
S(u,v)=\sum_{m,n=0}^\infty \frac{u^m\;v^n}{m!\;n!} S^{(m,n)}(0,0)
\ee
Likewise, consider the germ $[e]_v$ of $e$ 
\be \label{3.5}
e(t)=\sum_{n=0}^\infty \frac{t^n}{n!} e^{(n)}(0)
\ee
In order that $s_e e\subset S_{v,e}$ we just need to choose a 
parametrization 
of $S$ such that, say, $S(t,0)=e(t)$ which fixes the Taylor coefficients
\be \label{3.6}
S^{(m,0)}(0,0)=e^{(m)}(0)
\ee
for any $m$. By choosing the range of $t,u,v$ sufficiently small we can 
arrange that $s_e\subset S$.

We now choose the freedom in the remaining coefficients to 
satisfy the additional requirements. We must avoid that for finitely many,
say $N$, edges $e'_1,..,e'_n$ that there is any beginning segment 
$s_k$ of $e'_k$ with $s_k\subset S$. If $s_k$ would be contained in $S$
then there would exist an analytic function $t\mapsto v_k(t)$, 
such that $s_k(t)=S(t,v_k(t))$.  Notice that $v_k$ must be different
from the zero function 
in a sufficiently small neighborhood around $t=0$ as otherwise 
we would have $s_k=s_e$ which is not the case. For each $k$ let 
$n_k>0$ be the first derivative such that $v_k^{(n_k)}(0)\not=0$.
By relabeling the edges we may arrange that $n_1\le n_2\le ..\le n_N$.
Consider $k=1$ and take 
the $n_1-$th derivative at $t=0$. We find 
\be \label{3.7}
s_1^{(n_1)}(0)=S^{(n_1,0)}(0,0)+S^{(0,1)}(0,0) v_1^{(n_1)}(0)
\ee
Since $v_1^{(n_1)}(0)\not=0$ we can use 
the freedom in $S^{(0,1)}(0,0)$ in order to violate this equation.
Now consider $k=2$ and take the $(n_2+1)-$th derivative. We find 
\be \label{3.8}
s_2^{(n_2+1)}(0)=S^{(n_2+1,0)}(0,0)+2 S^{(1,1)}(0,0) v_2^{(n_2)}(0)+ 
S^{(0,1)}(0,0)v_2^{(n_2+1)}(0)
\ee
Since $v_2^{(n_2)}(0)\not=0$ we can use the freedom in $S^{(1,1)}$ in 
order to violate this equation. Proceeding this way we see that we can 
use the coefficients $S^{(k-1,1)}(0,0)$ in order to violate 
$s_k(t)=S(t,v_k(t))$ for $k=1,..,N$.

Having constructed the surfaces $S_{v,e}$ we can compute the associated 
vector field applied to a cylindrical function over $\gamma$
\be \label{3.9}
Y_j(S_{v,e}) p_\gamma^\ast f_\gamma=p_\gamma^\ast 
\sum_{e'\in E(\gamma)-\{e\},b(e')=v} \sigma(S_{v,e},e') R^j_{e'} f_\gamma
\ee
where by construction $|\sigma(S_{v,e},e')|=1$ for any 
$e'\not=e,\;b(e)=v$. Taking the commutator 
\be \label{3.10}
[Y_j(S_{v,e}),Y_k(S_{v,e})] p_\gamma^\ast f_\gamma=
f_{jkl}
p_\gamma^\ast 
\sum_{e'\in E(\gamma)-\{e\},b(e')=v} R^j_{e'} f_\gamma
\ee
Using the Cartan Killing metric normalization for the totally 
skew structure 
constants $f_{jkl} f_{lmj}=-\delta_{km}$ and writing 
\be \label{3.11}
R^j_v:=\sum_{e'\in E(\gamma),\;b(e')=v} R^j_{e'}
\ee
we get 
\be \label{3.12}
f_{jkl} [Y_k(S_{v,e}),Y_l(S_{v,e})] p_\gamma^\ast f_\gamma=
p_\gamma^\ast[R^j_v- R^j_e] f_\gamma 
\ee
Thus, if $n_v=|\{e\in E(\gamma);\;b(e)=v\}|$ denotes the valence of $v$
\ba \label{3.13}
Y^j_e p_\gamma^\ast f_\gamma &:=&
\{-f_{jkl} [Y_k(S_{v,e}),Y_l(S_{v,e})]+
\frac{1}{n_v-1} \sum_{e\in E(\gamma)} 
(f_{jkl} [Y_k(S_{v,e}),Y_l(S_{v,e})])\}p_\gamma^\ast f_\gamma
\nonumber\\
&=& p_\gamma^\ast R^j_e f_\gamma 
\ea
\\
Collecting the vector fields $Y^j_e$ for the Abelean and 
non-Abelean labels $j$ respectively and contracting 
them with $t^j_e$ and summing over $e\in E(\gamma)$
yields an appropriate vector field 
\be \label{3.14}
Y_\gamma(t_\gamma)=\sum_{e\in E(\gamma)} t^e_j Y^j_e
\ee
$\Box$\\

Lemma \ref{la3.1} has the following important implication: The algebra 
$\skripta$ also contains the vector field $Y_\gamma(t_\gamma)$ and therefore 
$\skriptw$ contains the corresponding Weyl element
$W_\gamma(t_\gamma)$.
Also, let us write $I_\gamma=(\{\pi_e\},\{m_e\},\{n_e\})_{e\in E(\gamma)}$
for a spin network $s=(\gamma, I_\gamma)$ over $\gamma$. Denoting
by $T_s=T_{\gamma,I_\gamma}$ the corresponding spin network function 
(where we also allow trivial $\pi_e$ for any $e$) we define for any two 
$\psi,\psi'\in {\cal H}_0$ the function
\be \label{3.16}
(t_\gamma,I_\gamma)\mapsto 
M_{\psi,\psi'}(t_\gamma,I_\gamma):=
\scpr{\psi}{T_{\gamma,I_\gamma}\;W_\gamma(t_\gamma) \psi'}_{{\cal H}_0}
\ee

We now exploit that for a compact connected Lie group the exponential map 
is onto. Thus, there exists a region $D_G\subset \Rl^{\dim(G)}$ such that
$\exp:\;D_G\to G;\;t\mapsto \exp(t^j\tau_j)$ is a bijection. 
Consider the measure $\mu$ on $D_G$ defined by $d\mu(t)=
d\mu_H(\exp(t^j\tau_j))$ where $\mu_H$ is the Haar measure on $G$. 
Finally, let 
$D_\gamma=\prod_{e\in E(\gamma)} D_G$ and let $L_\gamma$ be the space of 
the $I_\gamma$. We now define an inner product on the functions of 
the type (\ref{3.16}) by
\be \label{3.17}
(M_{\psi_1,\psi'_1},M_{\psi_2,\psi'_2})_\gamma:=
\int_{D_\gamma} d\mu(t_\gamma)\sum_{I_\gamma} 
\overline{M_{\psi_1,\psi'_1}(t_\gamma,I_\gamma)}\;\;
M_{\psi_2,\psi'_2}(t_\gamma,I_\gamma)
\ee
where $d\mu(t_\gamma)=\prod_{e\in E(\gamma)} d\mu(t_e)$. 

The inner product of the type (\ref{3.17}) is a crucial ingredient in an
elementary irreducibility proof of the Schr\"odinger representation of 
ordinary quantum mechanics (see for example \cite{22}) 
and we can essentially copy the 
corresponding argument. Of course, we must extend the proof somewhat in 
order to be able to deal with an infinite number of degrees of freedom. 
The following result prepares for that.
\begin{Lemma} \label{la3.2} ~~~~\\
i) For any $\psi_1,\psi_1',\psi_2,\psi_2'\in {\cal H}_0$ we have 
\be \label{3.18}
|(M_{\psi_1,\psi'_1},M_{\psi_2,\psi'_2})_\gamma|
\le 
||\psi_1||\;\;||\psi'_1||\;\;||\psi_2||\;\;||\psi'_2||.
\ee
ii) For any $\psi_1,\psi_1',\psi_2,\psi_2'\in {\cal H}_{0,\gamma}$ we have 
\be \label{3.19}
(M_{\psi_1,\psi'_1},M_{\psi_2,\psi'_2})_\gamma
=\scpr{\psi_2}{\psi_1}_{{\cal H}_0}\;\;\scpr{\psi'_1}{\psi'_2}_{{\cal H}_0}
\ee
where ${\cal H}_{0,\gamma}$ denotes the closure of the cylindrical 
functions over $\gamma$.
\end{Lemma}
Proof of lemma \ref{la3.2}:\\
We simply compute 
\ba \label{3.20}
&&(M_{\psi_1,\psi'_1},M_{\psi_2,\psi'_2})_\gamma
\nonumber\\
&=&
\int_{D_\gamma} d\mu(t_\gamma)\sum_{I_\gamma} 
\int_{\ab} d\mu_0(A)\int_{\ab} d\mu_0(A')
\overline{T_{\gamma,I_\gamma}(A)} T_{\gamma,I_\gamma}(A') 
\psi_1(A) \overline{[W_\gamma(t_\gamma)\psi'_1](A)}
\overline{\psi_2(A')} [W_\gamma(t_\gamma)\psi'_2](A')
\nonumber\\
&=&
\int_{D_\gamma} d\mu(t_\gamma)
\int_{\ab} d\mu_0(A)\int_{\ab} d\mu_0(A')
[\sum_{I_\gamma} \overline{T_{\gamma,I_\gamma}(A)} T_{\gamma,I_\gamma}(A')]
\psi_1(A) \overline{[W_\gamma(t_\gamma)\psi'_1](A)}
\overline{\psi_2(A')} [W_\gamma(t_\gamma)\psi'_2](A')
\nonumber\\
&=&
\int_{\ab} d\mu_0(A)\int_{\ab} d\mu_0(A')\int_{D_\gamma} d\mu(t_\gamma)
\delta_\gamma(A,A')
\psi_1(A) \overline{[W_\gamma(t_\gamma)\psi'_1](A)}
\overline{\psi_2(A')} [W_\gamma(t_\gamma)\psi'_2](A')
\ea
where we have defined the cylindrical $\delta-$distribution 
\be \label{3.21}
\delta_\gamma(A,A')=\prod_{e\in E(\gamma)} \delta_{\mu_H}(h_e[A],h_e[A'])
\ee
which arises due to the Plancherel formula 
\be \label{3.22}
\delta_{\mu_H}(g,g')=\sum_{\pi,m,n} \overline{T_{\pi,m,n}(g)}\;
T_{\pi,m,n}(g')
\ee
The interchange of integrals over $\ab\times \ab$ and the sum 
over $L_\gamma$ in (\ref{3.22}) is justified by the 
Plancherel theorem. 
i)\\
In order to evaluate the cylindrical $\delta-$distribution in (\ref{3.20})
we subdivide the degrees of freedom $A\in \ab$ into the set 
$\ab_\gamma=\ab_{|\gamma}$ and the complement 
$\ab_{\bar{\gamma}}=\ab-\ab_\gamma$ in the following sense: Each of the 
functions $f_1,f_1',f_2,f_2'$ is a countable linear combination of spin 
network functions $T_s$, each of which is cylindrical over some graph 
$\gamma(s)$. We may consider those functions as cylindrical over the graph 
$\gamma\cup\gamma(s)$ and since the edges $e\in E(\gamma)$ are 
holonomically independent, we can express each edge $\tilde{e}\in 
E(\gamma(s))$
as a finite composition of the edges of $E(\gamma)$ and some other edges 
$e'$ of $\gamma(s)\cup\gamma$ such that no segment of any of the 
$e'$ is a beginning segment of one of the $e$. 
Thus, each $T_s(A)$ depends on the 
$h_e,\;e\in E(\gamma)$ and some other $h_{e'}$ which are not finite 
compositions of the $h_e$. We can thus write symbolically 
for any $f\in {\cal H}_0$
\be \label{3.23}
f(A)=F(A_{|\bar{\gamma}},A_{|\gamma})
\ee
where the separation of the degrees of freedom is to be understood in the 
sense just discussed, that is, 
$A_{|\gamma}\in \ab_\gamma,\;A_{\bar{\gamma}}\in\ab_{\bar{\gamma}}$.
It just means that when expanding out inner products of $L_2$ functions 
into those of spin network functions, that one can perform the integrals
over the degrees of freedom in $\ab_\gamma$ and 
in $\ab{\bar{\gamma}}$ independently. Given a function 
of the type (\ref{3.23}) we define the measure on $\ab_\gamma$ by 
$\mu_{0\gamma}=\mu_0\circ p_\gamma^{-1}$ and the (effective) measure on 
$\ab_{\bar{\gamma}}$ by 
\begin{equation}
\label{3.24}
\begin{split}
&\int_{\ab_{\bar{\gamma}}} d\mu_{0\bar{\gamma}}(A_{|\bar{\gamma}})
[\int_{\ab_\gamma} d\mu_{0\gamma}(A_{|\gamma})
F(A_{|\bar{\gamma}},A_{|\gamma})]\cdot
\int_{\ab_\gamma} d\mu_{0\gamma}(A_{|\gamma})
[\int_{\ab_{\bar{\gamma}}} d\mu_{0\bar{\gamma}}(A_{|\bar{\gamma}})
F(A_{|\bar{\gamma}},A_{|\gamma})]\\
&\qquad:=\int_{\ab} d\mu_0(A) f(A)
\end{split}
\end{equation}
In order to perform concrete integrals of $f\in L_1(\ab,d\mu_0)$ 
over either $\ab_\gamma$ or $\ab_{\bar{\gamma}}$ we notice that 
all our occurring $f$ are countable linear combinations of spin network 
functions. Thus either integral can be written as a countable linear
combination of integrals over spin-network functions $T_s$ and then the 
prescription is to integrate only either over the degrees of freedom
$A(e),\;e\in E(\gamma)$ or $A(e'),\;e'\in 
E(\gamma(s)\cup\gamma)-E(\gamma)$ for each individual integral with
the corresponding product Haar measure.  
It follows that $\mu_0=\mu_{0\bar{\gamma}}\otimes \mu_{0\gamma}$ is a 
product measure. 

We may therefore 
neatly split (\ref{3.20}) as 
\begin{equation}
\label{3.25}
\begin{split}
(M_{\psi_1,\psi'_1},M_{\psi_2,\psi'_2})_\gamma
&=  
\int_{D_\gamma} d\mu(t_\gamma)
\int_{\ab_{\bar{\gamma}}} d\mu_{0\bar{\gamma}}(A_{|\bar{\gamma}})
\int_{\ab_{\bar{\gamma}}} d\mu_{0\bar{\gamma}}(A'_{|\bar{\gamma}})
\int_{\ab_\gamma} d\mu_{0\gamma}(A_{|\gamma})
\times\\
&\qquad \times
\Psi_1(A_{|\bar{\gamma}},A_\gamma) 
\overline{[W_\gamma(t_\gamma)\Psi'_1](A_{|\bar{\gamma}},A_\gamma)}
\overline{\Psi_2(A'_{|\bar{\gamma}},A_\gamma)} 
[W_\gamma(t_\gamma)\Psi'_2](A'_{|\bar{\gamma}},A_\gamma)
\end{split}
\end{equation}
In order to evaluate the Weyl operators, consider a spin network function 
$T_s$ cylindrical over $\gamma(s)$ which we write in the form 
\be \label{3.26}
T_s(A)=F(\{h_{e'}\}_{e'\in E(\gamma\cup\gamma(s))-E(\gamma)},
\{h_e\}_{e\in E(\gamma)})
\ee
Our concrete vector field $Y_\gamma(t_\gamma)$ involves a finite 
collection of surfaces to which the edges $e\in E(\gamma)$ are already 
adapted in the sense that they are all of a definite type 
(``in'', ``out'', ``up'' or ``down'') and we may w.l.g. assume that the 
same is true for the $e'$. Then it is easy to see that the action 
of $Y_\gamma(t_\gamma)$ on $T_s$ is given by
\be \label{3.27a}
Y_\gamma(t_\gamma) T_s=p_{\gamma(s)\cup\gamma}^\ast
[\sum_{e'\in E(\gamma\cup\gamma(s))-E(\gamma)}
t^{e'}_j(t_\gamma) R^j_{e'}+\sum_{e\in E(\gamma)} t^e_j R^j_e] 
F
\ee
where $t^{e'}_j(t_\gamma)$ is a certain linear combination of the 
$t^e_j$ depending on $e'$ and the concrete surfaces $S_e,S_{v,e}$ used in 
the construction of $Y_\gamma(t_\gamma)$.
Since the beginning segments of the $e',e$ are mutually independent,
the corresponding vector fields commute and it follows that 
\ba \label{3.27}
(W_\gamma(t_\gamma) T_s)(A) &=&
F(\{e^{t^{e'}_j(t_\gamma) \tau_j}
h_{e'}\}_{e'\in E(\gamma\cup\gamma(s))-E(\gamma)},
\{e^{t^e_j \tau_j} h_e\}_{e\in E(\gamma)})
\nonumber\\
&=& F(\{W_\gamma(t_\gamma)h_{e'}
W_\gamma(t_\gamma)^{-1}\}_{e'\in E(\gamma\cup\gamma(s))-E(\gamma)},
\{W_\gamma(t_\gamma) h_e W_\gamma(t_\gamma)^{-1}\}_{e\in E(\gamma)})
\ea

Consider now any $L_2$ function $\psi$. Since it is a countable linear 
combination of spin network functions we can generalize
(\ref{3.27}) to
\be \label{3.28}
(W_\gamma(t_\gamma)\psi)(A)
=\psi(W_\gamma(t_\gamma)A_{|\bar{\gamma}}W_\gamma(t_\gamma)^{-1},  
W_\gamma(t_\gamma)A_{|\gamma}W_\gamma(t_\gamma)^{-1})
\ee
where the crucial point is that for each $t_\gamma\in D_\gamma$ 
the map $\alpha_{t\gamma}:\;\ab\to \ab;\;A\mapsto W_\gamma(t_\gamma) A 
W_\gamma(t_\gamma)$
is just some right or left translation. We can thus estimate (notice that
we can interchange the sequence of integration w.r.t. the factors of a 
product measure)
\begin{align*} 
|(M_{\psi_1,\psi'_1},M_{\psi_2,\psi'_2})_\gamma|
&\le
\int_{D_\gamma} d\mu(t_\gamma)\int_{\ab_\gamma} d\mu_{0\gamma}(A_{|\gamma})
\times\\
&\qquad\times
[\int_{\ab_{\bar{\gamma}}} d\mu_{0\bar{\gamma}}(A_{|\bar{\gamma}})
|\Psi_1(A_{|\bar{\gamma}},A_\gamma)|\;\;
|\Psi'_1(\alpha_{t_\gamma}(A_{|\bar{\gamma}}),
\alpha_{t_\gamma}(A_\gamma))|
\times\\
&\qquad \times 
[\int_{\ab_{\bar{\gamma}}} d\mu_{0\bar{\gamma}}(A'_{|\bar{\gamma}})
|\Psi_2(A'_{|\bar{\gamma}},A_\gamma)|\;\; 
|\Psi'_2(\alpha_{t_\gamma}(A'_{|\bar{\gamma}}),\alpha_{t_\gamma}(A_\gamma))|
\\
&\le
\int_{D_\gamma} d\mu(t_\gamma)\int_{\ab_\gamma} d\mu_{0\gamma}(A_{|\gamma})
||\Psi_1(A_\gamma)||_{|\bar{\gamma}}\;\;
||\Psi'_1(\alpha_{t_\gamma}(A_\gamma))||_{\bar{\gamma}}
||\Psi_2(A_\gamma)||_{\bar{\gamma}} \;\;
||\Psi'_2(\alpha_{t_\gamma}(A_\gamma))||_{\bar{\gamma}}
\end{align*}
where we have used the Cauchy Schwarz inequality applied to functions
such as $\Psi_1(A_\gamma)$   
on $L_2(\ab_{\bar{\gamma}},d\mu_{0\bar{\gamma}})$ defined by 
$[\Psi_1(A_\gamma)](A_{|\bar{\gamma}})=\Psi_1(A_{|\bar{\gamma}},A_\gamma)$.
Here it was crucial to note that due to the bi-invariance of the 
measure $\mu_{0\bar{\gamma}}$ we have e.g.
\begin{equation*}
\int_{\ab_{\bar{\gamma}}} d\mu_{0\bar{\gamma}}(A_{|\bar{\gamma}})
|\Psi'_1(\alpha_{t_\gamma}(A_{|\bar{\gamma}}),
\alpha_{t_\gamma}(A_\gamma))|^2
=\int_{\ab_{\bar{\gamma}}} d\mu_{0\bar{\gamma}}(A_{|\bar{\gamma}})
|\Psi'_1(A_{|\bar{\gamma}},
\alpha_{t_\gamma}(A_\gamma))|^2
=||\Psi'_1(\alpha_{t_\gamma}(A_\gamma))||^2_{\bar{\gamma}}
\end{equation*}
To see this, expand $\psi'_1$ into spin-network functions. Then 
the integral is of the form 
\begin{align*}
\sum_{m,n=1}^\infty \bar{z}_m z_n
&\int_{\ab_{\bar{\gamma}}} d\mu_{0\bar{\gamma}}(A_{|\bar{\gamma}})
\overline{T_{s_m}(\alpha_{t_\gamma}(A))}\, T_{s_n}(\alpha_{t_\gamma}(A))\\
&=\sum_{m,n=1}^\infty \bar{z}_m z_n
\int_{G^{|E(\gamma(s_m)\cup\gamma(s_n)\cup\gamma)-E(\gamma)|}}\;\;
[\prod_{e'\in E(\gamma(s_m)\cup\gamma(s_n)\cup\gamma)-E(\gamma)}
d\mu_H(h_{e'})]\;\;\times\\
&\qquad\qquad\times
\overline{T_{s_m}(\{e^{t^{e'}_j(t_\gamma)\tau_j}h_{e'}\},
\{e^{t^e_j\tau_j}h_e\})}\;\; 
T_{s_n}(\{e^{t^{e'}_j(t_\gamma)\tau_j}h_{e'}\},
\{e^{t^e_j\tau_j}h_e\})\\
&=\sum_{m,n=1}^\infty \bar{z}_m z_n
\int_{G^{|E(\gamma(s_m)\cup\gamma(s_n)\cup\gamma)-E(\gamma)|}}\;\;
[\prod_{e'\in E(\gamma(s_m)\cup\gamma(s_n)\cup\gamma)-E(\gamma)}
d\mu_H(h_{e'})] 
\times\\
&\qquad\qquad\times
\overline{T_{s_m}(\{h_{e'}\},\{e^{t^e_j\tau_j}h_e\})}\;\; 
T_{s_n}(\{h_{e'}\},\{e^{t^e_j\tau_j}h_e\})\\
&=\sum_{m,n=1}^\infty \bar{z}_m z_n
\int_{\ab_{\bar{\gamma}}} d\mu_{0\bar{\gamma}}(A_{|\bar{\gamma}})
\overline{T_{s_m}(A_{|\bar{\gamma}},\alpha_{t_\gamma}(A_{|\gamma}))} \;\;
T_{s_m}(A_{|\bar{\gamma}},\alpha_{t_\gamma}(A_{|\gamma}))\\
&=
\int_{\ab_{\bar{\gamma}}} d\mu_{0\bar{\gamma}}(A_{|\bar{\gamma}})
|\Psi'_1(A_{|\bar{\gamma}},\alpha_{t_\gamma}(A_{|\gamma}))|^2
\end{align*}
We now exploit that
\be \label{3.31}
\alpha_{t_\gamma}(A_{|\gamma})
=\{e^{t^e_j\tau_j}h_e\}_{e\in E(\gamma)}
\ee
and introduce new integration variables $h'_e:=g(t_e)h_e$ where
$g(t_e)=\exp(t^e_j\tau_j)$. Since by definition 
\be \label{3.32}
d\mu(t_\gamma)=\prod_{e\in E(\gamma)} d\mu(t_e) 
=\prod_{e\in E(\gamma)} d\mu_H(g(t_e)) 
\ee
we can estimate further
\begin{align*}
|(M_{\psi_1,\psi'_1},M_{\psi_2,\psi'_2})_\gamma|
&\le
\int_{G^{|E(\gamma)|}} \prod_{e\in E(\gamma)} d\mu_H(g_e)
\int_{\ab_\gamma} d\mu_{0\gamma}(A_{|\gamma})
\times\\
&\qquad\qquad\times
||\Psi_1(A_{|\gamma})||_{|\bar{\gamma}}\;\;
||\Psi'_1(\{g_e A(e)\}_{e\in E(\gamma)})||_{\bar{\gamma}} 
||\Psi_2(A_{|\gamma})||_{\bar{\gamma}} \;\;
||\Psi'_2(\{g_e A(e)\}_{e\in E(\gamma)})||_{\bar{\gamma}}
\\
&=
[\int_{\ab_\gamma} d\mu_{0\gamma}(A_{|\gamma})
||\Psi_1(A_{|\gamma})||_{|\bar{\gamma}}\;\;
||\Psi_2(A_{|\gamma})||_{\bar{\gamma}}]
[\int_{\ab_\gamma} d\mu_{0\gamma}(A'_{|\gamma})
||\Psi'_1(A'_{|\gamma})||_{\bar{\gamma}}\;\;
||\Psi'_2(A'_{|\gamma})||_{\bar{\gamma}}]
\\
&\le
||\;\;||\Psi_1||_{\bar{\gamma}}\;\;||_\gamma\;\;
||\;\;||\Psi'_1||_{\bar{\gamma}}\;\;||_\gamma\;\;
||\;\;||\Psi_2||_{\bar{\gamma}}\;\;||_\gamma\;\;
||\;\;||\Psi'_2||_{\bar{\gamma}}\;\;||_\gamma
\end{align*}
where we have used the Fubini theorem and have again applied the Cauchy 
Schwarz inequality to functions in $L_2(\ab_\gamma,d\mu_{0\gamma})$.
But
\begin{align*}
||\;\;||\Psi_1||_{\bar{\gamma}}\;\;||^2_\gamma
&=\int_{\ab_\gamma} d\mu_{0\gamma}(A_{|\gamma})
|\;\;||\Psi_1(A_{|\gamma})||_{\bar{\gamma}}\;\;|^2\\
&=
\int_{\ab_\gamma} d\mu_{0\gamma}(A_{|\gamma})
\int_{\ab_{\bar{\gamma}}} d\mu_{0\bar{\gamma}}(A_{|\bar{\gamma}})
|\Psi_1(A_{|\bar{\gamma}},A_{|\gamma})|^2
=\int_{\ab} d\mu_0(A)|\psi_1(A)|^2=||\psi_1||^2_{{\cal H}_0}
\end{align*}
so we get (\ref{3.18}).\\
\\
ii)\\
If all functions in question are cylindrical $L_2-$functions over $\gamma$
then the integrals over $\ab_{|\bar{\gamma}}$ are trivial and 
(\ref{3.25}) simplifies to
\begin{align*}
(M_{\psi_1,\psi'_1},M_{\psi_2,\psi'_2})_\gamma
&=
\int_{D_\gamma} d\mu(t_\gamma)
\int_{\ab_\gamma} d\mu_{0\gamma}(A_{|\gamma})
\Psi_1(A_\gamma) 
\overline{[W_\gamma(t_\gamma)\Psi'_1](A_\gamma)}
\overline{\Psi_2(A_\gamma)} 
[W_\gamma(t_\gamma)\Psi'_2](A_\gamma)\\
&=
\int_{\ab_\gamma} d\mu_{0\gamma}(A_{|\gamma})
\int_{\ab_\gamma} d\mu_{0\gamma}(A'_{|\gamma})
\Psi_1(A_\gamma) 
\overline{\Psi'_1(A'_\gamma)}
\overline{\Psi_2(A_\gamma)} 
\Psi'_2(A'_\gamma)\\
&=
[\int_{\ab} d\mu_0(A) \overline{\psi_2(A)} \psi_1(A)]\;\;
[\int_{\ab} d\mu_0(A') \overline{\psi'_1(A')} \psi'_2(A')]
\\
&=
\scpr{\psi_2}{\psi_1}_{{\cal H}_0}\;\;
\scpr{\psi'_1}{\psi'_2}_{{\cal H}_0}
\end{align*}
that is, (\ref{3.19}).\\
$\Box$\\
\\
Proof of theorem \ref{th3.1}:\\
Suppose that the representation $\pi_0$ of $\skriptw$ is not 
irreducible, that is, not every vector is cyclic. Thus, we find 
non zero vectors $\psi,\psi'\in {\cal H}_0$ such that 
\be \label{3.36}
\scpr{\psi}{a\psi'}=0\;\; \forall \;\; a\in\skriptw
\ee
Since the cylindrical functions lie dense in ${\cal H}_0$, for any
$\epsilon>0$ we find a graph $\gamma$ and functions $f,f'$ cylindrical
over $\gamma$ such that
\be \label{3.37}
||\psi-f||<\epsilon,\;\;||\psi'-f'||<\epsilon
\ee
From (\ref{3.36}) we have in particular that 
$M_{\psi,\psi'}(t_\gamma,I_\gamma)=0$ for all 
$t_\gamma\in D_\gamma,\;I_\gamma\in L_\gamma$, hence
\ba \label{3.38}
0 &=& (M_{\psi,\psi'},M_{\psi,\psi'})_\gamma
\\
&=& 
(M_{\psi-f,\psi'},M_{\psi,\psi'})_\gamma
+(M_{f,\psi'-f'},M_{\psi,\psi'})_\gamma
+(M_{f,f'},M_{\psi-f,\psi'})_\gamma
+(M_{f,f'},M_{f,\psi'-f'})_\gamma
+(M_{f,f'},M_{f,f'})_\gamma
\nonumber\\
&=& 
(M_{\psi-f,\psi'},M_{\psi,\psi'})_\gamma
+(M_{f,\psi'-f'},M_{\psi,\psi'})_\gamma
+(M_{f,f'},M_{\psi-f,\psi'})_\gamma
+(M_{f,f'},M_{f,\psi'-f'})_\gamma
+||f||^2\;\;||f'||^2
\nonumber
\ea
where (\ref{3.19}) has been used. Exploiting $\psi,\psi'\neq 0$ we may choose 
$\epsilon<||\psi||,||\psi'||$ and using (\ref{3.37}) and (\ref{3.18})
we have 
\ba \label{3.39}
&&(||\psi||-\epsilon)^2\; (||\psi'||-\epsilon)^2
\\
&\le& ||f||^2\;\;||f'||^2
\nonumber\\
&\le& 
||\psi-f||\;\;||\psi'|| \;\;||\psi||\;\;||\psi'||
+||f||\;\;||\psi'-f'||\;||\psi||\;\;||\psi'||
\nonumber\\
&& +||f||\;\;||f'||\;||\psi-f||\;\;||\psi'||
+||f||\;\;||f'||\;\;||f||\;\;||\psi'-f'||
\nonumber\\
&\le& 
\epsilon\{||\psi'||^2 \;\;||\psi||
+(||\psi||+\epsilon)\;||\psi||\;\;||\psi'||
+(||\psi||+\epsilon)\;\;(||\psi'||+\epsilon)\;\;||\psi'||
+(||\psi||+\epsilon)^2\;\;(||\psi'||+\epsilon)\}\nonumber
\ea
Since this inequality holds for all $\epsilon$ we can take $\epsilon\to 0$
and find 
\be \label{3.40}
||\psi||^2\;\;||\psi'||^2=0
\ee
that is, either $\psi=0$ or $\psi'=0$ in contradiction to our assumption.
Hence $\pi_0$ is irreducible.$\qquad\Box$

\section{Conclusions}
\label{s4}

In this paper we have shown that the Ashtekar -- Isham -- Lewandowski
representation $\pi_0$ of the Weyl algebra $\skriptw$ 
underlying diffeomorphism 
invariant quantum gauge field theories for compact gauge groups is 
irreducible. While this has been common belief, this is, to the best of 
our knowledge, the first time that this has been shown rigorously.

From a mathematical point of view it can be considered as an extension
of the known irreducibility proofs for the Weyl algebras of quantum scalar
field theories to the non-Abelean context. 

From a physical point of view,
irreducibility is an important concept because it makes the superselection 
structure of the theory trivial: There are no distinguished sectors in the 
Hilbert space that are left invariant and one does not need to worry about 
the charges that distinguishes those sectors. It has been known that 
the vector $1\in {\cal H}_0$ is cyclic already for the subalgebra Cyl of
$\skriptw$ consisting of cylindrical functions. However, that does not
imply that the representation is irreducible because if there would be a 
non-trivial, closed, invariant subspace $V$ then, since $\skriptw$ 
is closed under involution, its orthogonal complement
$V^\perp$ is also closed and invariant and all we knew is that $1$ could 
be uniquely decomposed as $1=P_V\cdot 1\oplus P_{V^\perp}\cdot 1$ where 
$P_V,\;P_{V^\perp}$ are the associated orthogonal projections.

If a theory has non-trivial closed invariant subspaces then this 
is typically a sign for the fact that either the algebra $\skriptw$ is 
too small
(it has no elements that map between the sectors) or that the 
representation space ${\cal H}_0$ is too large because interesting physics 
can already be captured by one of its invariant subspaces
\cite{23}. In this paper
we have shown that this is not the case for $\pi_0$, thus giving yet
one more piece of evidence for the physical assumption that it is a
suitable kinematical starting point for the quantization of diffeomorphism 
invariant gauge field theories, which seems to be suitable in order to
define the quantum (constraint) dynamics of Quantum General Relativity
(coupled to all known matter) \cite{24}.\\ 
\\
{\large Acknowledgments}\\
\\
H.S. thanks the Albert-Einstein-Institut for hospitality. This work was 
supported in part by NSF grant PHY-0090091

\end{document}